\documentclass[transmag]{IEEEtran}
\usepackage{latexsym}
\usepackage{cite}
\usepackage{mciteplus}
\usepackage{amsthm}
\usepackage{graphicx}
\usepackage{float}
\usepackage{subfigure}
\usepackage{url}
\usepackage{stfloats}
\usepackage{amsmath}
\usepackage{amsfonts}
\usepackage{amssymb}
\usepackage{eqnarray}
\usepackage{subeqnarray}
\usepackage{array}
\usepackage{flushend}
\usepackage{setspace}
\usepackage{paralist}
\usepackage{suffix}
\usepackage{mathtools}
\usepackage{subfig}
\usepackage{epsfig}
\usepackage{psfrag}
\usepackage{epstopdf}
\usepackage{multirow}
\usepackage{tabularx}
\usepackage{textcomp}
\usepackage{xcolor}
\usepackage{longtable}
\def\BibTeX{{\rm B\kern-.05em{\sc i\kern-.025em b}\kern-.08em T\kern-.1667em\lower.7ex\hbox{E}\kern-.125emX}}

\begin{document}
\title{Smart and Secure Wireless Communications via 
Reflecting Intelligent Surfaces: A Short Survey}
\author{Abdullateef Almohamad, Anas M. Tahir, Ayman Al-Kababji, Haji M. Furqan, Tamer Khattab, \IEEEmembership{Senior Member, IEEE},\\ Mazen O. Hasna, \IEEEmembership{Senior Member, IEEE}, and Huseyin Arslan, \IEEEmembership{Fellow, IEEE}
\thanks{This publication was made possible by the grants NPRP12S-0225-190152 and GSRA6-2-0521-19034 from the Qatar National Research Fund (a member of The Qatar Foundation). The statements made herein are solely the responsibility of the author[s].}
\thanks{A. Almohamad, A. M. Tahir, A. Al-Kababji, T. Khattab, and M. Hasna are with the Department of Electrical Engineering, Qatar University, Doha, Qatar (e-mails: \{abdullateef, anas.tahir, ayman.alkababji\}@ieee.org, \{tkhattab, hasna\}@qu.edu.qa).}
\thanks{H. M. Furqan is with the Department of Electrical and Electronics Engineering, Istanbul Medipol University, Istanbul, 34810, Turkey (e-mail: hamadni@medipol.edu.tr).}
\thanks{H. Arslan is with the Department of Electrical and Electronics Engineering, Istanbul Medipol University, Istanbul, 34810, Turkey and with the Department of Electrical Engineering, University of South Florida, Tampa, FL, 33620, USA (e-mail: huseyinarslan@medipol.edu.tr).}
}

\IEEEtitleabstractindextext{\begin{abstract}
With the emergence of the internet of things (IoT) technology, wireless connectivity should be more ubiquitous than ever. In fact, the availability of wireless connection everywhere comes with security threats that, unfortunately, cannot be handled by conventional cryptographic solutions alone, especially in heterogeneous and decentralized future wireless networks. In general, physical layer security (PLS) helps in bridging this gap by taking advantage of the fading propagation channel. Moreover, the adoption of reconfigurable intelligent surfaces (RIS) in wireless networks makes the PLS techniques more efficient by involving the channel into the design loop. In this paper, we conduct a comprehensive literature review on the RIS-assisted PLS for future wireless communications. We start by introducing the basic concepts of RISs and their different applications in wireless communication networks and the most common PLS performance metrics. Then, we focus on the review and classification of RIS-assisted PLS applications, exhibiting multiple scenarios, system models, objectives, and methodologies. In fact, most of the works in this field formulate an optimization problem to maximize the secrecy rate (SR) or secrecy capacity (SC) at a legitimate user by jointly optimizing the beamformer at the transmitter and the RIS's coefficients, while the differences are in the adopted methodology to optimally/sub-optimally approach the solution. We finalize this survey by presenting some insightful recommendations and suggesting open problems for future research extensions.
\end{abstract}

\begin{IEEEkeywords}
Physical Layer Security, PLS, Reconfigurable Intelligent Surface, RIS, Secrecy Outage Probability, Secrecy Rate.
\end{IEEEkeywords}}

\maketitle

\section{Introduction}\label{sec1}
\IEEEPARstart{D}{ue} to the considerable increase in the number of wirelessly communicating devices, different innovative technologies have been proposed in the literature to enhance the energy and spectrum efficiency along with the reliability and security of wireless communication systems. The future applications from 5G wireless communication's perspective include three use cases with diverse requirements such as ultra-reliable low latency communication (URLLC), enhanced mobile broadband (eMBB), and massive machine-type communication (mMTC). The promising physical layer technologies to fulfill the requirements of the above-mentioned applications include cognitive radio (CR), cooperative communication, massive multiple-input multiple-output (ma-MIMO), millimeter-wave, orthogonal frequency division multiplexing (OFDM) numerologies, and so on \cite{8796365}.  

{The future wireless networks are expected to support highly (energy and spectral) efficient, secure, reliable, and flexible design for emerging applications of 6G and beyond \cite{8782879}. In order to achieve this goal, rigorous efforts have been undertaken in the research and development of wireless communications. However, overall progress has been relatively slow. This is due to the fact that conventional wireless communication designers have focused only on transmitter and receiver ends while considering the wireless communication environment as an uncontrollable factor. Moreover, it is also presumed that this factor has usually a negative effect on communication efficiency and reliability, and consequently, needs to be compensated.}

{Recently, reconfigurable intelligent surfaces (RIS) received focused attention due to their significant capability in enabling a smart and controllable wireless propagation environment \cite{8910627}. Specifically, an RIS is a uniform planar array that consists of low-cost passive reflecting elements. Each element in an RIS can be controlled to smartly adjust the amplitude and/or phase of incoming electromagnetic waves, thus, rendering the direction and strength of the wave highly controllable at the receivers. This feature can be exploited to add different signals constructively/destructively to enhance/weaken their overall strength at different receivers. Thus, RISs can be used to enhance the signal-to-noise ratio (SNR), data rate, security, and/or the coverage probability. In \cite{wu2019intelligent}, the problem of minimizing the transmit power in the RIS-assisted MISO system under quality of service (QoS) constraints is investigated. Specifically, it is revealed that a squared power gain in terms of the number of reflecting elements can be achieved by applying active and passive beamforming. In addition, RIS-assisted systems offer significantly higher power-efficient alternatives to conventional multi-antenna amplify-and-forward relaying systems \cite{huang2019reconfigurable}. Moreover, the employment of real-time tunable RIS can be used to mitigate and/or eliminate the multipath and Doppler effects caused by the movement of the mobile receiver/transmitter \cite{basar2019reconfigurable}.}

Motivated by the appealing advantages, RIS-assisted networks have been investigated in many different contexts such as capacity and rate improvement analysis \cite{hu2018capacity,jung2020uplink}, power efficiency optimization \cite{huang2019reconfigurable,fu2019intelligent}, communication reliability \cite{jung2019reliability,chafii2016necessary}, {physical layer security (PLS), and so on. PLS has emerged as a powerful complementary solution for enhancing the security of future wireless communication systems besides cryptographic algorithms. These approaches exploit the dynamic characteristics of the wireless channel such as channel randomness, interference, noise, fading, dispersion, diversity, separability, reciprocity, etc., to ensure secure communication \cite{8509094, 9018254}. Due to RIS's capability in enabling a smart controllable wireless propagation, it is a promising solution to enhance the performance of PLS techniques, even for a challenging scenario when PLS techniques are ineffective. More specifically, when the legitimate node and illegitimate node are in the same direction, many PLS techniques including conventional beamforming, directional modulation, artificial noise (AN), etc., cannot fully ensure secure communication. However, this issue can be addressed with the employment of RIS near to legitimate/eavesdropping user along with the proper design of beamforming to enhance/weaken the signal strength at the legitimate/eavesdropping user, thus, significantly enhancing the overall security of the system.} {In fact, the efficiency of RISs in supporting a wireless communication system in terms of secrecy, for instance, is driven by some practical limitations. Ideally, an RIS can be seen as a continuous surface of reflecting elements (zero-spaced elements) with continuous induced phase shifts and reflection coefficients by each element. However, practically speaking, the controllable reflecting elements can be achieved using mechanical actuation, special materials (e.g., graphene), and electronic devices (e.g., positive-intrinsic-negative (PIN) diodes) \cite{wu2020intelligent}. Thus, it has a response switching time, frequency and angle-of-arrival (AoA) dependent response, and inter-element coupling effects. Furthermore, practically, the induced phase shift is non-linearly coupled with the reflection coefficient \cite{abeywickrama2020intelligent}, hence, optimizing the RIS beamforming/reflection should involve both the phase shifts and the reflection coefficients.}

{The major advantage of having an RIS in a communication system is the ability to perform passive beamforming, which is done at a middle point in the channel, unlike the traditional active beamforming at the base station (BS) side. This extra degree of freedom has been proved to enhance the system performance in terms of multiple metrics, especially in terms of PLS which is completely dependent on the system's ability to accurately direct the signal beam into a desired path (or exclude it). Moreover, with the aid of an RIS, the coverage area can be, more than ever, tailored as per the network designer requirements. Furthermore, with the passive intelligent reflection, the noise at the reflected signal is not amplified as with the conventional relays.}

{The adoption of RISs comes with an increased system complexity. For example, in a PLS application, conventionally the active beamforming is optimized to support the system secrecy, while with the presence of an RIS in the loop, joint optimization is required to take advantage of the passive beamforming which is highly dependent on the quality of the acquired CSI. This actually leads to another challenge, which is the acquisition of the CSI itself at the RIS's side, taking into account the high number of the RIS’s reflecting elements and their passive nature. Moreover, the channel reciprocity assumption in time-division duplex (TDD) channels, which simplifies the channel estimation process, is no longer valid with the presence of an RIS in the system \cite{wu2020intelligent}. Additionally, under the far-field propagation assumption, the communication channel through an RIS suffers from double path loss, known as a product-distance path loss model, which needs to be compensated for either in the link budget or by increasing the number of reflecting elements \cite{wu2020intelligent}.}

An increasing number of recent works have studied RIS-assisted communications from a PLS point of view. In general, there are two main research directions under the PLS concept, namely, information-theoretic secrecy and covert communications. The former focuses on improving the secrecy rate (SR) of legitimate users by exploiting the dynamic features of wireless communications, for example, random channel, fading, interference, and noise, etc., to prevent the eavesdropper from decoding leaked data while ensuring that the legitimate user can decode it successfully. The covert communications direction, on the other hand, considers hiding the existence of communication from being detected by an enemy \cite{lu2019intelligent,8509094}. In this survey, we will focus on the first direction, and for simplicity, it will be referred to as PLS. For more details on PLS in general, we refer the reader to our comprehensive survey in \cite{8509094}.

\begin{figure*}
    \centering
    \includegraphics[trim={1.1in 0.5in 0.9in 0.3in},clip,width=1\linewidth]{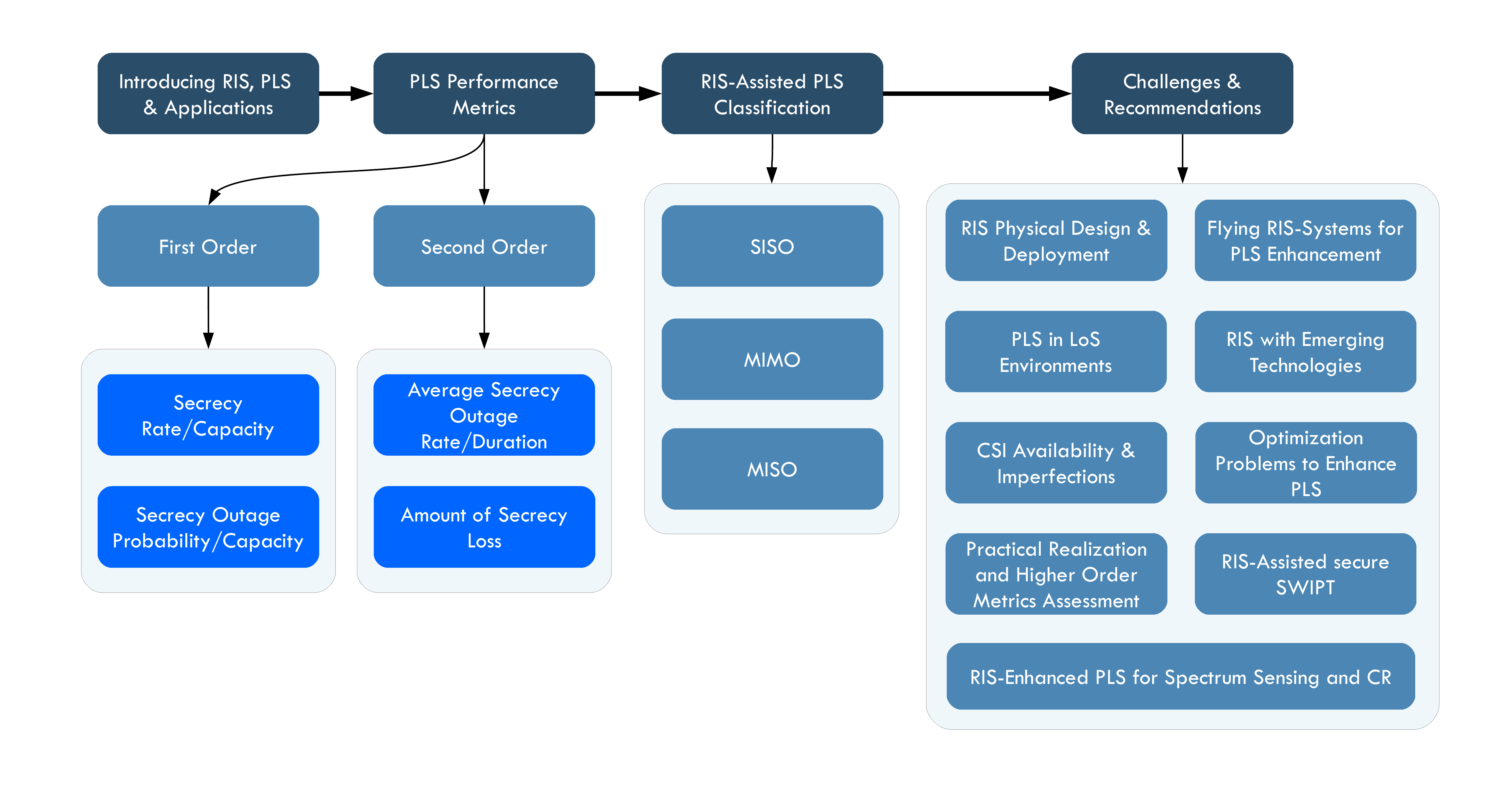}
    \caption{A diagram to show the structure of the paper.}
    \label{fig:paper_structure}
\end{figure*}

In this survey, and to the best of our knowledge, all related papers to the RIS-assisted PLS in wireless networks are systematically reviewed. Some common shortcomings in the current literature which lead to open extensions for research are highlighted. The outline and structure of this survey are shown in Fig.~\ref{fig:paper_structure}.

The remainder of this paper is organized as follows, the most common secrecy performance metrics are presented in Section \ref{sec2}. Then, categorization of the RIS-assisted PLS studies is included in Section \ref{sec3}. Recommendations and open research directions are listed in Section \ref{sec4}. Finally, concluding remarks are drawn in Section \ref{sec5}.


\section{Secrecy Performance Metrics}\label{sec2}
In this section, we present a brief but comprehensive review of the most commonly used metrics for assessing PLS. We include both first and second-order metrics.
\subsection{Secrecy Rate/Capacity (SR/SC)}
SR is one of the fundamental metrics to measure the secrecy of a communication system. It represents the amount of information in bits per second that can be securely delivered to the receiver over a given channel. Specifically, the achievable SR is the difference between the achievable data rate on the legitimate and the eavesdropper channels, respectively, which is given as
\begin{equation}
R_s = \max\{R_D - R_E,0\},
\end{equation}
where $R_D$ and $R_E$ represent the achievable rates over the legitimate user and the eavesdropper channels, respectively. Practically, positive SR can be achieved by active, at the transmitter, and/or passive, at the RIS, beamforming by degrading the eavesdropper channel while improving the legitimate user one.

Similar to Shannon channel capacity, the SC is defined as the upper bound of the SR \cite{liang2009information}. The SC of the Wyner degraded wiretap channel is given as \cite{wyner1975wire}
\begin{equation}
C_s = \sup_{p(X)}\{I(X;Y) - I(X;Z)\},
\end{equation}
where $I(\cdot,\cdot)$ represents the mutual information, $X$ and $Y$ represent the input and the output of the legitimate user channel, respectively, $Z$ denotes the output of the eavesdropper channel and $p(X)$ is the input probability distribution. The SC, for a given channel realization, can be written in terms of Shannon's channel capacities of the legitimate user and the eavesdropper as follows \cite{leung1978gaussian}
\begin{equation}
\label{SC}
C_s = \max\{\log_2(1+\gamma_D) - \log_2(1+\gamma_E) ,0 \}, 
\end{equation}
where $\gamma_D$ and $\gamma_E$ represent the instantaneous SNR at the legitimate user and the eavesdropper, respectively. The ergodic capacity is obtained by averaging \eqref{SC} as per the available channel state information (CSI).

\subsection{Secrecy Outage Probability/Capacity (SOP/SOC)}
Similar to the known outage probability in communication systems, the secrecy outage probability (SOP) is defined as the probability of the event when the instantaneous SC falls below a given target SR, which is written as follows
\begin{equation}\label{SOP1}
\text{SOP}(R_{\text{th}}) = \text{Pr} \{C_s < R_{\text{th}}\}.
\end{equation}

In fact, the definition in \eqref{SOP1} does not differentiate between the outage due to non reliable legitimate channel and the outage due the leakage of information to the eavesdropper. Therefore, another more explicit definition is proposed in \cite{bai2014outage} as follows
\begin{equation}\label{SOP2}
    \text{SOP}(R_{\text{code}},R_{\text{th}}) = 1-\text{Pr} \{C_D \geq R_{\text{code}}, C_s \geq R_{\text{th}}\},
\end{equation}
where $C_D$ is the instantaneous legitimate channel capacity, $R_{\text{code}}$ is the coding rate of the transmitted message. It is clear, in this definition, that the secrecy outage event happens when the coding rate $R_{\text{code}}$ fails to satisfy Shannon's reliable transmission condition in addition to having the SC below a target SR threshold $R_{\text{th}}$. 

A related and widely adopted metric is the secrecy outage capacity (SOC), which is defined as the maximum achievable SC, $C_{\text{out}}$, that guarantees an SOP of less than a threshold $\epsilon$ \cite{gungor2013secrecy}, which is expressed as follows
\begin{equation}
   \max \{C_{\text{out}}\} \text{ with } \text{Pr}\{C_s < C_{\text{out}}\} = \epsilon
\end{equation}

\subsection{Average Secrecy Outage Rate/Duration (ASOR/ASOD)}
The aforementioned performance metrics are based on the first-order statistics, however, incorporating the second-order statistics in the secrecy performance metrics offers a better understanding of the dynamics of the performance.
Two secrecy performance metrics that fall under this category were proposed in \cite{omri2018average}. Namely, the average secrecy outage rate (ASOR) and the average secrecy outage duration (ASOD). The former, ASOR denoted by $\mathcal{R}(R_{\text{th}})$, measures the SC's average rate of crossing a given threshold level $R_{\text{th}}$, whereas the ASOD measures, in seconds, the average duration in which the system remains in a secrecy outage status. ASOD is expressed, at a given threshold $R_{\text{th}}$, in terms of the SOP and the ASOR as follows
\begin{equation}
    \mathcal{T}(R_{\text{th}}) = \frac{\text{SOP}(R_{\text{th}})}{\mathcal{R}(R_{\text{th}})}.
\end{equation}

\subsection{Amount of Secrecy Loss (ASL)}
Recently, the authors in \cite{li2020amount}, proposed a new PLS performance metric, the amount of secrecy loss (ASL), based on the second order statistics of the SC. The ASL measures the amount of information leakage to the eavesdropper, which is expressed as
\begin{equation}
    \text{ASL} = \frac{\text{E}\{C_s^2\}}{\text{E}\{C_s\}^2}-1,
\end{equation}
where $\text{E}\{\cdot\}$ is the statistical expectation operator.

\section{Categorizing Recent Studies on Wireless RIS-Reinforced Secrecy}\label{sec3}
In this section, we classify the most recent works on RIS-reinforced PLS in wireless communications in terms of the considered system model. As we noted, most of the related works are aimed to maximize the SR/SC while the differences were found in the considered system model and the methodology to optimize the objective in hand. The classification is done based on the number of antennas at the transmitter and the receiver. In what follows, we interchangeably use the terms Alice/BS, Bob/legitimate and Eve/eavesdropper.
\begin{figure}
\centering
\begin{tabular}{c}
\includegraphics[trim={0.7in 0in 0.7in 0in},clip,width=1\linewidth]{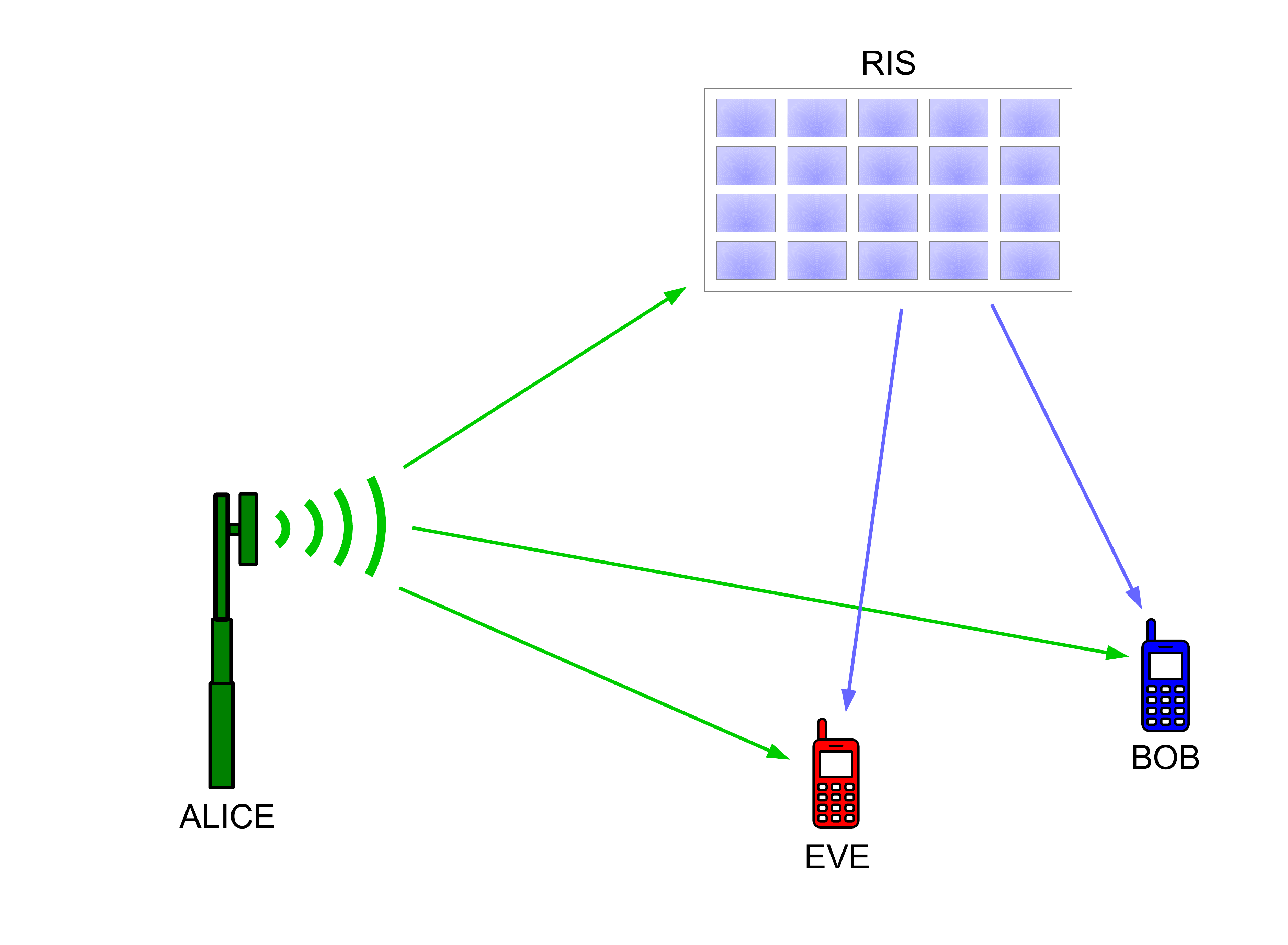}\\
(a) \\
\includegraphics[trim={0.7in 0in 0.7in 0in},clip,width=1\linewidth]{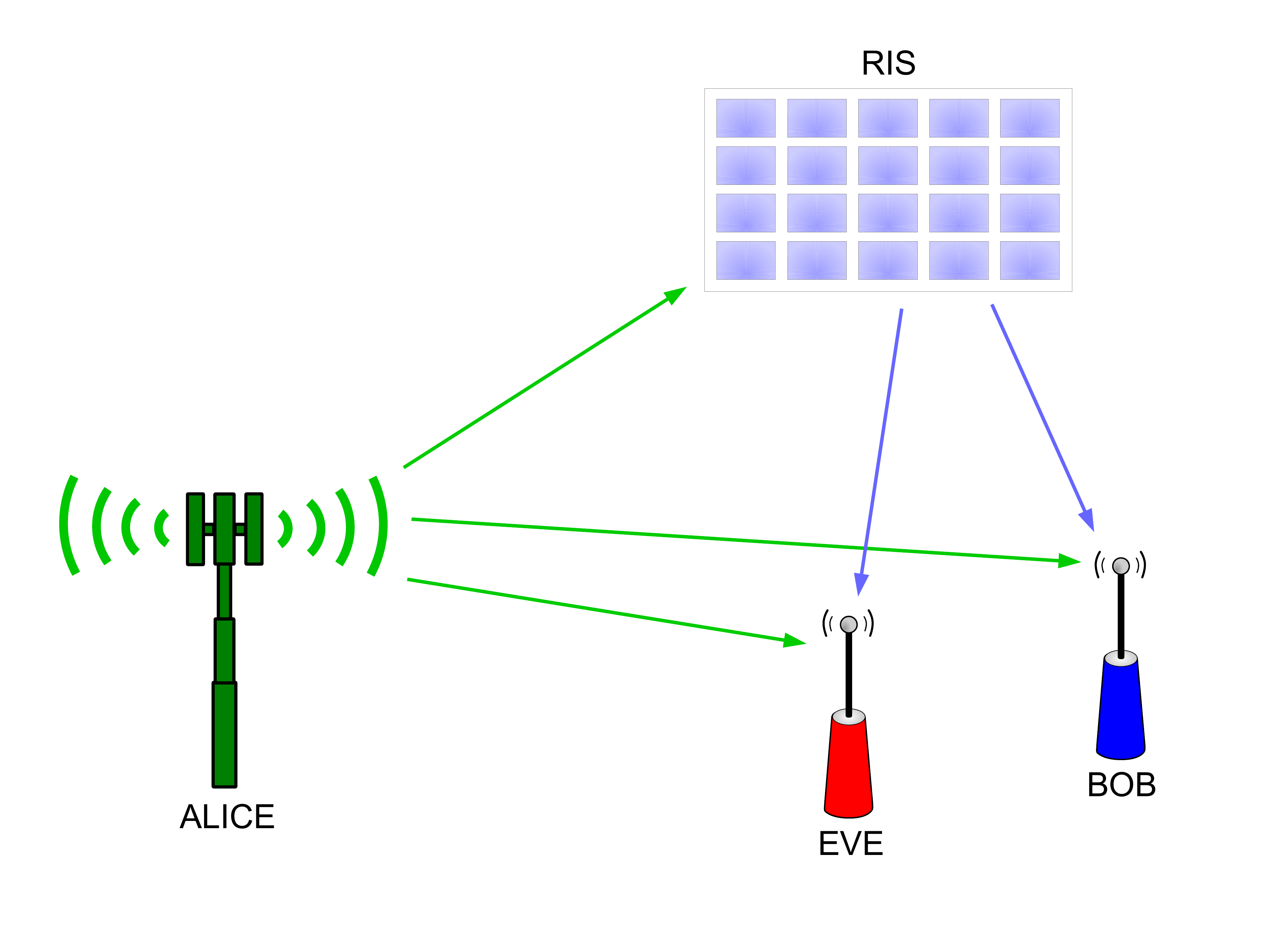}\\
(b)
\end{tabular}
\caption{\small{Most Common System Models in the Literature (a) SISO, (b) MIMO}}
\label{fig1}
\end{figure}

\subsection{SISO System Model}
The simplest setup we encountered in the literature assumes a single-antenna transmitter, Alice, willing to securely deliver a message to a single-antenna legitimate user, Bob, in the presence of a single-antenna eavesdropper, Eve, as shown in Fig.~\ref{fig1}-(a).

Yang et al. \cite{Yang2020} studies the secrecy performance of an RIS-assisted SISO communication link in the presence of a line-of-sight (LoS) links between the RIS and the eavesdropper and the legitimate user. Single RIS is considered with $N$ reflecting elements placed between the source and the legitimate user. The CSI of the legitimate user is assumed to be known at the RIS. Thus, the RIS can induce the required phase shifts on the reflected signal to maximize the received SNR at the legitimate user. The analytical expression of the SOP is derived as an evaluation metric to assess the secrecy performance. The analytical and simulation results show that the presence of an RIS significantly enhances the SR and the enhancement is driven by the number of RIS's reflecting elements. However, the secrecy performance slightly drops when the eavesdropper enjoys a LoS link with the RIS as well. This is due to optimizing the RIS's induced phases to maximize the SNR at the legitimate user but ignoring the effect it exercises on the eavesdropper's received SNR. In \cite{long2020reflections}, an unmanned aerial vehicle (UAV) equipped with an RIS is used as a mobile relay between a group of users and a BS. The authors focus on the maximization of secrecy energy efficiency by joint optimization of the passive beamforming, the user-UAV association, the UAV trajectory, and the transmit power. Alternating optimization (AO) and successive convex approximation (SCA) algorithms are used, where the objective is to attain fairness in SR among users and minimum energy.

In vehicular ad hoc networks (VANET), PLS is a major concern, due to the broadcast nature of the wireless channels. Many papers, \cite{makarfi2019physical,ai2018physical,xu2019physical}, have considered the analysis of PLS under such high mobility conditions and dynamic environments. In fact, RIS is proven to help compensating the multipath and Doppler effects in wireless propagation channels \cite{basar2019reconfigurable}. Capitalizing on this advantage, an analytical approach is followed in \cite{makarfi2020physical,makarfi2020reconfigurable} to optimize the SC in a VANET, where two setups are proposed to investigate the PLS. The first setup assumes a source, a destination, and an eavesdropping vehicle communicating with the support of an RIS mounted on a nearby building, while the second setup assumes that the source vehicle has an RIS coupled with its transmitter. A double Rayleigh distribution is assumed between the mobile ends, and the Meijer G-function is used to obtain the probability distribution function (PDF) of the received source, which slightly complicates the analysis. The reported results are similar to those in \cite{Yang2020}. Furthermore, the authors study the effect of varying the number of RIS's reflecting elements and the distance between the source and the RIS. Specifically, as the number of reflecting elements increases, the SR/SC improves because better beamforming can be achieved, and the SR/SC degrades while increasing the source-RIS distance which is due to fading and path loss effects.
A recent work, by the authors, \cite{OptimizingSec2020}, investigates the effectiveness of RIS-assisted network by introducing a weighted variant of the SC definition. Simulation results show that the existence of a reliable LoS link dominates the system’s SC. However, it can be further enhanced by optimizing the RIS-induced phase shifts. In addition, it is shown that the RIS-assisted system with non-line-of-sight (NLoS) links achieves comparable SC to that of dominant LoS link systems with unknown RIS-Eve CSI.

{A general indoor system model is considered in \cite{hong2020secrecy}, where a SISO system with multiple Bobs and Eves is investigated. Analytical based genetic algorithm (GA) is utilized to find the optimal tile-allocation-and-phase-shift-adjustment (TAaPSA) strategy for the RIS to optimize the average SR. The obtained results show that the number of Eves in the system has a significant effect on optimal trend of TAaPSA strategy. For low number of Eves, the SR can be maximized by simultaneously enhancing average rate of Bobs and degrading the average rate of the Eves. In the contrary case, the RIS can be fully utilized to boost the average rate of Bob. }

\subsection{MIMO System Model}
Dong and Wang in \cite{Dong2020} consider a MIMO system model, as shown in Fig.~\ref{fig1}-(b), where the BS, the eavesdropper, and the legitimate user have multiple antennas. However, the system has NLoS transmission. The objective again is to maximize the SR at the legitimate user by jointly optimizing the transmit covariance matrix and the phase shift matrix of the RIS's reflecting elements. Solving this non-convex problem is intractable, hence, an AO algorithm is proposed assuming complete knowledge of CSI of both the legitimate user, and the eavesdropper at the RIS and the BS. The proposed solution is shown to monotonically converge within a number of iterations that is dependent on the number of antennas at the BS, legitimate user, and eavesdropper. On the other hand, the authors in \cite{Hong2020} opt for including the LoS transmission channel and using AN, consequently, rendering the objective function more challenging to solve. The proposed optimization algorithm is block coordinate descent (BCD) aided by the majorization minimization (MM) algorithm. The results show how increasing the number of RIS’s elements can increase the SR at the expense of burdening the optimization algorithm with a larger phase shift matrix to optimize.

Similarly, the authors in  \cite{jiang2020intelligent} considered a LoS channel as in \cite{Hong2020} with the same objective, but the authors consider the case of discrete phase shifts at the RIS after solving the optimization problem under the continuous phases assumption. As we know that the optimization problem under the continuous phases is non-convex, it can be solved using an AO method, where for a given RIS reflect coefficients, SCA is used to optimize the transmit covariance matrix. Next, for a given transmit covariance matrix, the AO method is used again to optimize the individual elements' phase shift of the RIS one by one, given the other elements' shifts at each step. Numerical simulations show that 3-bits quantized phase shifts yields an acceptable SR with negligible loss as compared to the continuous phase shifts case. Noting the large scale of this MIMO setup, it is clear that the adopted AO method in solving the optimization problem suffers from high computational complexity especially when we consider a large scale RIS and a high number of antennas at the BS, the legitimate user, and the eavesdropper. Furthermore, the optimization of the phases matrix and the covariance matrix are independent problems only if the BS-RIS channel is a rank-one matrix \cite{Feng2019}, thus, the AO gives a sub-optimal solution in the full rank channel case.

{In order to highlight the benefits of having an RIS supporting the secure communications with an RIS-empowered eavesdropper, the authors in \cite{alexandropoulos2020safeguarding} consider an eavesdropper with a supporting passive eavesdropping RIS that competes against a legitimate RIS to impair the system's secrecy. As expected, it has been shown that a non-zero SR cannot be achieved with AN and preceding with the absence of the legitimate RIS. However, a legitimate RIS with $L$ reflecting elements can safeguard secure communication against a larger than $3L$-elements eavesdropping RIS.}

\begin{figure}
\centering
\begin{tabular}{c}
\includegraphics[trim={0.7in 0in 0.7in 0in},clip,width=1\linewidth]{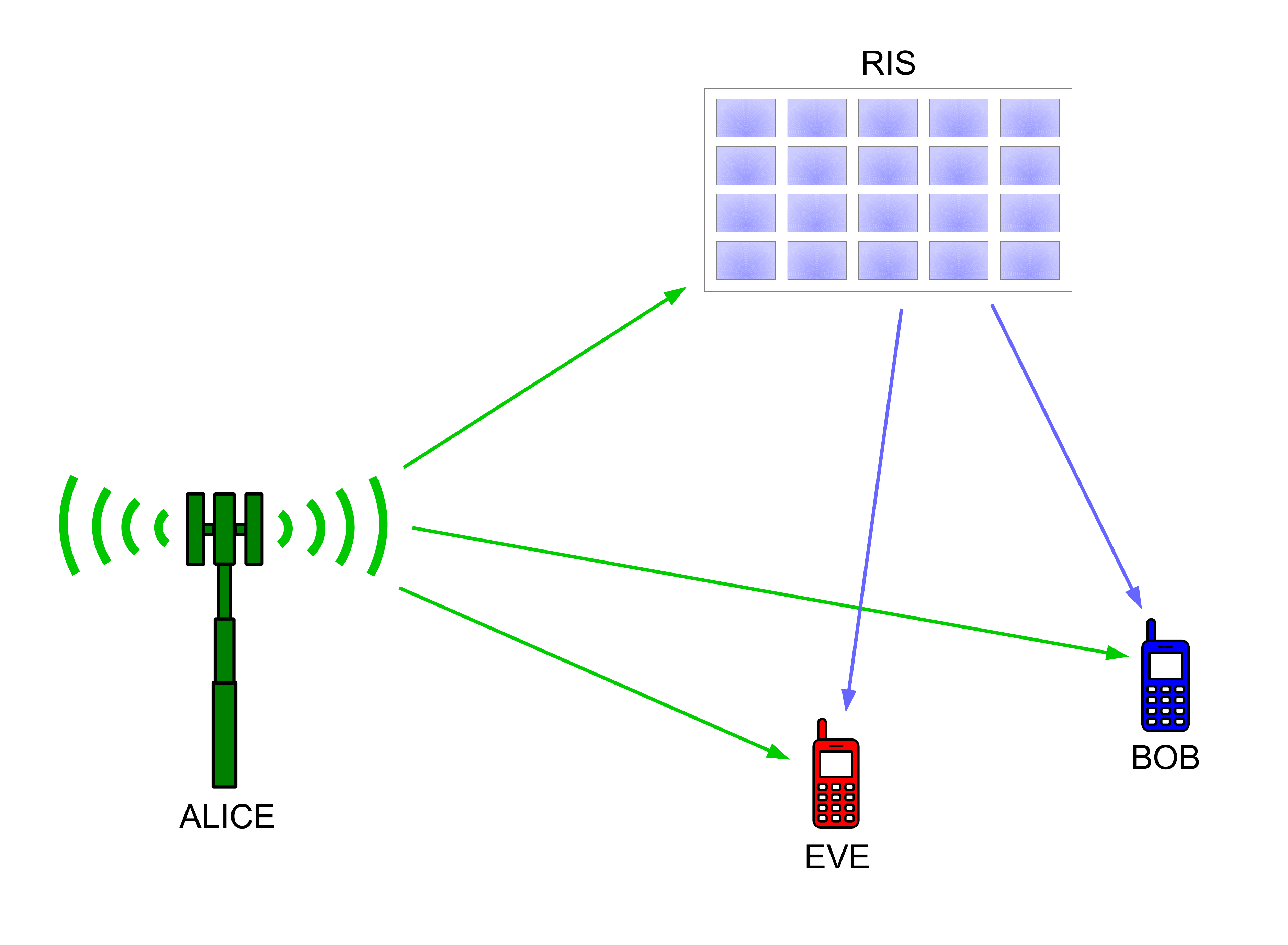}\\
(a)\\
\includegraphics[trim={0.7in 0in 0.7in 0in},clip,width=1\linewidth]{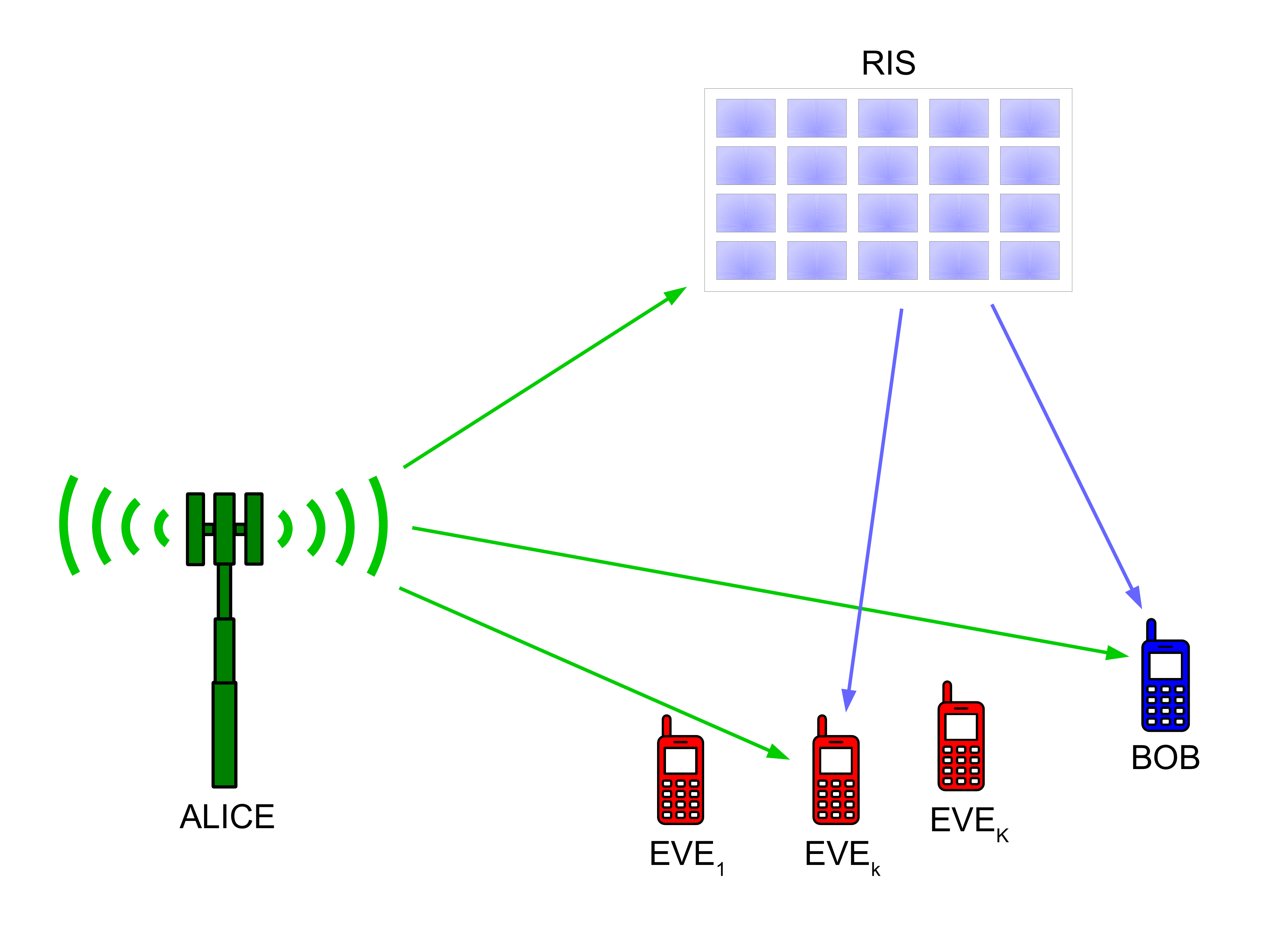}\\
(b)
\end{tabular}
\caption{\small{Most Common System Models in the literature (a) MISO with single eavesdropper, (b) MISO with multiple eavesdroppers}}
\label{fig2}
\end{figure}

\subsection{MISO System Model}
As most of the relevant works fall under this subsection, we further classify them based on the following: the number of users in the system, the CSI availability/acquiring assumptions, the pursued methodology, and practicality-related assumptions.
\vspace{-0.5em}
\subsubsection{Single Bob and Eve}
Many works, \cite{Feng2019,chensecrecy,shi2019enhanced,Yu,Shen2019,Cui2019,ning2019improving,song2020truly}, considered a simplified system model as shown in Fig.~\ref{fig2}-(a). Where a multiple antennas BS, is considered, communicating a secure messages to a single user (Bob) in the presence of a single Eve, both having a single antenna.
In \cite{chensecrecy,Yu,Shen2019,Cui2019}, optimization problems are proposed to maximize the SR at the legitimate user by jointly optimizing the beamforming at the BS and the phase shifts at the RIS. Due to the intractability of this problem, different methodologies have been adopted. In \cite{chensecrecy}, RIS discrete phase shifts is assumed and an AO method is followed, where for a given RIS phase shifts matrix, the optimal BS precoder is obtained using Rayleigh-Ritz theorem. On the other hand, for a given BS precoder matrix, a cross-entropy-based algorithm is adopted to optimize the RIS phase shifts matrix. In \cite{Yu}, two efficient joint optimization techniques are proposed, namely: AO-MM and BCD. It is revealed that the AO-MM algorithm is favorable for large-scale RIS-assisted systems, while the BCD is superior for wireless systems with small-scale RISs. In addition, the obtained results show that installing a large scale RIS yields better enhancement in terms of SR and is more energy-efficient as compared to enlarging the transmit antenna array size. However, continuous phase shifts is assumed at the RIS which serves as an upper bound on the practical performance. An AO-based algorithm is developed in \cite{Shen2019} to solve the joint problem of optimizing the transmit covariance matrix of the BS and the RIS's phase shift matrix providing closed-form and semi-closed-form solutions, respectively. Moreover, the obtained AO-based solution, with the help of fractional optimization (FO), is extended to to the case where the eavesdropper can have multiple antennas. It is shown that the SR degrades when the number of eavesdropper antennas increases, as it starts to achieve higher rates. 

It is worth noting that the simulated distance-dependent scenarios conducted by different works considered one-dimensional (1D) movements only where the transmitter, the legitimate user, and the eavesdropper lie on a planar surface. However, no work addressed the two-dimensional (2D)/three-dimensional (3D) movements which are the case in many emerging scenarios. A more practical scenario is considered in \cite{Cui2019}, where the channels of the legitimate user and the eavesdropper are spatially correlated, and the latter has a stronger channel. An AO-SDR based method is adopted and similar results are achieved as in \cite{Shen2019}. Similarly, the authors in \cite{ning2019improving} considered the joint optimization problem but for the Terahertz (THz) communications scenarios. An AO-based algorithm is developed, where for a given beamforming matrix, the optimal RIS's phase shifts is obtained by leveraging the characteristics of the THz channel. Then, given the obtained phase shifts, the optimal beamformer is derived by utilizing the Rayleigh-Ritz theorem \cite{golub1996cf}. It is worth mentioning here that the RIS in this work operates in two modes, namely: the sensing and reflection modes. In the former, the RIS reflecting elements turn into active antennas supported with RF chains. However, this assumption is not power efficient noting the high number of reflecting elements at the RIS.

Energy efficiency is another constraint that is incorporated in \cite{Feng2019,shi2019enhanced}. In \cite{Feng2019}, the energy consumption is investigated assuming NLoS links between the BS and legitimate user/eavesdropper. Thus, the authors aimed for minimizing the consumed energy in BS-RIS link through beamforming at the transmitter and optimizing the phase shifts at the RIS. It is shown that the beamforming and the phase matrix optimization problems are independent in the rank-one channel case. However, in the full-rank channel case, they are inseparable. For both channel rank cases, projected gradient descent (PGD) and semidefinite relaxation (SDR) are used to solve the joint optimization problem. The performance of both optimization algorithms is analyzed to find out that both yield similar results, however, the SDR algorithm converges faster than the PGD. {Within a related context, RIS-supported simultaneous wireless information and power transfer (SWIPT) is investigated in \cite{shi2019enhanced}. A MISO system supported by an RIS is considered to improve the delivered energy to an energy harvesting receiver (EHR) in addition to information transfer to a legitimate receiver with the presence of an eavesdropper. An AO method is adopted after relaxing the non-convex problem using the SDR technique. The resulting high complexity algorithm is further improved to reduce the complexity using the SCA approximation. The reported results reveal that, with the presence of an RIS, the harvested power can be doubled under SR constraint as compared to the traditional case with no RIS. A CR system is considered in \cite{xiao2020intelligent}, where the legitimate user, Bob, is considered as a secondary user that is trying to access a licensed band in the presence of an eavesdropper. A nearby RIS is installed to support a secure communication for Bob while maintaining an upper bound on the interference level at the primary user. An AO algorithm is proposed to maximize the SR of Bob subject to total power constrain at Alice, and inference power constrain at the primary Bob in the presence of Eve. Simulation results show that SR can be significantly enhanced compared with no RIS case. In addition, even with the constraint on the interference, the SR keeps increasing with transmit power unlike the case with no RIS which shows a saturation in the SR at higher transmit power levels.}

{Different from the ideal assumption in existing literature that full Eve’ CSI is available, Wang et al. \cite{wang2020intelligent} consider a more practical scenario where Eve’s CSI is not available.  To enhance the system security given a total transmit power at Alice, two joint beamforming and jamming optimization algorithms are proposed based on OM and MM methods. The transmit power is firstly optimized at Alice to meet the QoS at Bob, while AN is emitted to jam Eve by using the residual power at Alice. The AN is projected onto the null space of Bob's channel to ensure that only Eve is jammed. As compared to the ideal case of full CSI, security could still be guaranteed by relaxing the QoS threshold at Bob as well as increasing the number of RIS reflecting elements. Unlike the current research trend on utilizing RIS to enhance the system secrecy, lyu et al. \cite{lyu2020irs} propose the use of an RIS as a passive jammer to attack legitimate communication between BS and Bob. BCD, SDR, and Gaussian randomization techniques are utilized to jointly optimize reflection coefficient magnitudes and discrete phase shifts at the RIS to diminish the signal-to-interference-plus-noise ratio (SINR) at Bob. Obtained results show that the performance of the proposed RIS-based jammer outperforms that of conventional active jamming attacks in some scenarios, especially when the distance between RIS and Bob is small (less than 10m).} Noting the high computational complexity of solving the joint optimization problem, discussed so far, the authors in \cite{song2020truly} introduce a machine learning (ML) model by utilizing deep neural network (DNN) to maximize the system's SR in real time. Simulation results show that the DNN can achieve comparable results to the conventional optimization methods with simpler and faster implementation.
\vspace{0.5em}
\subsubsection{Multiple Bobs or Eves}
Fewer works, \cite{Wang2019,Guan2020,chu2019intelligent}, considered the case when multiple eavesdroppers are attacking a single legitimate user as shown in Fig.~\ref{fig2}-(b). Jamming is a common technique to deny the eavesdropper from receiving any useful signal, which can be done even actively as in \cite{wang2020intelligent,Wang2019,Guan2020} or passively with the aid of an illegitimate RIS as in \cite{lyu2020irs}. Wang et al. \cite{Wang2019}, considered the energy efficiency by optimizing the beamforming at the transmitter, as in \cite{chensecrecy, Feng2019,wang2020intelligent}, aided by a cooperative jamming with the support of an optimized RIS phase shifts. An SDR-based energy-efficiency maximization problem is defined to optimize the transmit power, the independent cooperative jamming, and the RIS reflection coefficients under a constraint on the SR. The proposed scheme is highly energy-efficient even with high jamming power, which implies the significance of the cooperative jammer that is introduced. Moreover, it also outperforms the other reported schemes in maintaining a high SR along with being energy efficient. On the other hand, Guan et al. \cite{Guan2020} investigate the RIS's effectiveness in improving the SR, with an AN induced by the transmitter to jam the eavesdroppers instead of the cooperative jammer in \cite{Wang2019}. The objective is to maximize the SR at the legitimate user by jointly optimizing the transmitter beamforming, the AN (jamming), and the passive beamforming at the RIS. An AO-based method is used to optimize the three dependent elements in the objective function. It is shown that the use of AN jamming requires fewer reflecting elements on the RIS in order to maintain a specific SR threshold. In a related context, where the power efficiency is considered, the authors in \cite{chu2019intelligent}, study the role of RIS in minimizing the transmitted power while maintaining a secrecy level. Considering the non-convex nature of the problem, an AO algorithm and an SDR method to optimize the secure and power-efficient transmission are proposed. Few simulation scenarios are presented illustrating the distance effect on the transmitted power showing that as the legitimate user moves away from the BS and getting closer to the RIS, the overall system energy efficiency increases as less power needs to be transmitted. In addition, in scenarios where the eavesdropper is close to BS/RIS, higher transmitted power is needed to keep a secure communication. Unlike the above mentioned works, multiple legitimate users in the presence of a single eavesdropper is considered in \cite{Xu2019}, where the beamformer, the AN's covariance matrix at the BS and the RIS's phase shifts are optimized to maximize the average sum of the legitimate users' SRs. A sub-optimal solution is obtained for the non-convex problem using AO method, and applying SCA, SDR and manifold optimization (MO) approaches. Results show an increase in the average SR with the support of the RIS and the AN generated at the BS. However, the analysis is limited to the case with NLoS which simplifies the problem in hand.
\vspace{0.5em}
\subsubsection{Multiple Bobs and Eves}
On the other hand, as a more realistic scenario, multiple legitimate users are assumed to be attacked by multiple eavesdroppers in \cite{Yu2019,yang2020deep,yang2020intelligent,chen2019intelligent}. Considering the high computational complexity of the involved optimization problem for this generalized scenario, ML-based algorithms would provide a fast and flexible solutions. For instance, {Yang et al. \cite{yang2020deep} consider a novel deep reinforcement learning (RL) approach to achieve optimal beamforming policy in a dynamic environment. In addition, post-decision state (PDS) and prioritized experience replay (PER) schemes are employed to enhance the secrecy performance and learning efficiency. Simulation results show that the proposed algorithm outperforms conventional optimization approaches by achieving a higher average SR per user and higher QoS satisfaction probabilities. Another critical security challenge against legitimate transmission is malicious jamming launched by smart jammers. Authors in \cite{yang2020intelligent} consider the use of RIS to mitigate the jamming interference and enhance the communication performance by proposing joint optimization approach using fuzzy win or learn fast-policy hill-climbing method. The fuzzy model helps to estimate the dynamic jamming model, where uncertain environments states are represented as aggregate of fuzzy states. Simulation results show that the proposed approach can improve both transmission protection level and RIS-assisted system rate compared with existing solutions.}

An extension to the work in \cite{Xu2019} is made in \cite{Yu2019} to include multiple legitimate users and multiple eavesdroppers with two RISs instead of one, rendering the problem to be much more challenging. {Moreover, the potential eavesdroppers are roaming users within another BS, where their feedback leaked signal to their corresponding BS can be utilized for coarse CSI estimation at the BS. To improve the estimation, a deterministic model is adopted to characterize the CSI's uncertainty.} The work is extended to account for the SOP along with the average SR. The significant contribution is in incorporating several RISs with a uniform number of elements instead of having a single RIS with a huge number of elements. However, the LoS analysis is not included, which is an important aspect in practical scenarios. The authors in \cite{chen2019intelligent} propose an optimization problem to maximize the minimum SR at the legitimate users by jointly optimizing the BS beamforming and the RIS's phase shifts. Furthermore, the reflecting elements of the RIS are assumed to be discrete, and a spatial channel correlation between the legitimate users and the eavesdropper is assumed as well. Again, due to the non-convexity of the problem, and hence the non-tractability, an AO method and a path-following algorithm are proposed to maximize the objective function.

Based on the conducted survey, we assemble the reviewed works in Table~\ref{table1} which summarizes the different assumptions and system models with methodologies and performance metrics in the conducted literature review. Moreover, Table~\ref{table2} classifies them in terms of the system model (SISO, MISO, and MIMO) and the adopted methodology to approach the considered objective.

\begin{table*}[ht!]
\renewcommand{\arraystretch}{1.2}
\centering
\caption{\small{Summary of the different assumptions, system models, methodologies, and performance metrics. Where $N_a$, $N_b$, $N_e$ and $N_r$ denote the number of antennas at Alice, Bob, Eve and RIS, respectively, $K$, $L$ and $M$ denote the number of Eves, Bobs and RISs, respectively, $I_i$ denotes the number of iterations of the $i^{\textrm{th}}$ loop, and $I_\epsilon$ denotes the total number of iterations to achieve the target accuracy $\epsilon$.}}
\begin{tabular}
{
|>{\centering\arraybackslash}m{0.8cm}
|>{\centering\arraybackslash}m{2.2cm}
|>{\centering\arraybackslash}m{2.5cm}
|>{\centering\arraybackslash}m{3cm}
|>{\centering\arraybackslash}m{4.8cm}
|>{\centering\arraybackslash}m{2.1cm}
|
}
\hline
\textbf{Ref.} & 
\textbf{System Model} & 
\textbf{Methodology} &
\textbf{Assumptions} &
\textbf{Complexity} &
\textbf{Metric} \\ \hline 

\cite{Yang2020} &
  SISO &
  Analytical &
  Quasi-static flat fading &
  -- &
  SOP \\ \hline
  \cite{long2020reflections} &
  SISO, Bob=L &
  Approximation (AO-SCA) &
  Rayleigh/Rician fading &
  -- &
  Secrecy Energy Efficiency \\ \hline
  \cite{makarfi2019physical,makarfi2020reconfigurable} &
  SISO &
  Analytical &
  Double Rayleigh fading &
  -- &
  Average SC \& SOP \\ \hline
  \cite{OptimizingSec2020} &
  SISO &
  Analytical &
  Rayleigh flat fading &
  -- &
  SC \\ \hline 
\cite{hong2020secrecy} &
SISO, Eve=K, Bob=L &
Analytical (GA) &
Ray model, Rice distribution &
-- &
SOP \\ \hline
\cite{Dong2020} &
  MIMO &
  Approximation (AO-MM) &
  Gaussian wiretap &
  -- &
  SR \\ \hline 
\cite{Hong2020} &
  MIMO &
  Approximation (BCD-MM)  &
  Narrow-band \& non-dispersive channel &
  -- &
  SR \\ \hline 
\cite{jiang2020intelligent} &
  MIMO, Eve=K &
  Approximation (AO-SCA) &
  Discrete phase shifts &
  $\begin{multlined}
  O(I_3(I_1(N_e^3+N_a^6)\ + \\I_2N_r(N_b^3+N_e^3-\log_2\epsilon))) \end{multlined}$ &
  
  SR \\ \hline
\cite{Feng2019} &
  MISO &
  Approximation (PGD-SDR)&
  Multiple channel models&
  -- &
  Average SR \\ \hline
\cite{chensecrecy} &
  MISO &
  Approximation (AO) &
  Discrete phase shifts &
  -- &
  SR \\ \hline
\cite{shi2019enhanced} &
  MISO &
  Approximation (AO-SDR-SCA)  &
  Rayleigh fading, EHR &
  $\begin{gathered}
  O(N_a^8 + N_r^8)\,\, \textrm{\&}\,\, O(N_a^3 + N_r^3) \\
  \textrm{AO-SDR \& AO-SCA}
  \end{gathered}$ &
  SR \& Harvested Power\\ \hline
\cite{Yu} &
  MISO &
  Approximation (AO-MM-BCD) &
  Rayleigh fading &
  -- &
  Average SR \\ \hline
  \cite{Shen2019} &
  MISO &
  Approximation (AO-FO) &
  Rayleigh fading &
  -- &
  SR \\ \hline
\cite{Cui2019} &
  MISO &
  Approximation (AO-SDR) &
  Bob \& Eve channels are correlated &
  $\begin{multlined}
  O(I_1(N_a^3+(N_r+1)^{3.5})) 
  \end{multlined}$&
  Average SR \\ \hline
  \cite{ning2019improving} &
  MISO &
  Approximation (AO-FO) &
  Saleh Valenzuela model &
  -- &
  SR \\ \hline
 \cite{song2020truly} &
  MISO &
  Approximation (DNN) &
  Quasi-static flat-fading &
  -- &
  Average SR \\ \hline
\cite{xiao2020intelligent} &
  MISO, Primary Receiver=1  &
  Approximation (AO) &
  RIS-assisted Gaussian CR wiretap channel &
O($N_a^3$+$N_a^2$)  \,\,
or \,\,
O($N_a^3+I_{\epsilon}N_a^2$)  &
  SR \\ \hline
\cite{wang2020intelligent} &
  MISO  &
  Approximation (OM and MM) &
  Quasi-static flat-fading,  Unknown Eve's CSI &
  O($I_1N_a^2$) &
  SR \\ \hline
\cite{Wang2019} &
  MISO, Eve=K &
  Approximation (AO-SDR-FO) &
  Quasi-static flat fading &
  -- &
  SR \\ \hline
\cite{Guan2020} &
  MISO, Eve=K &
  Approximation (AO) &
  Quasi-static flat-fading &
  $\begin{multlined}
  O(I_3(I_1\max\{N_r,K\}^4N_r^{0.5}\\
  +I_2\max\{N_a,K\}^4N_a^{0.5})) 
  \end{multlined}$ &
  SR \\ \hline
  \multirow{2}{*}{\cite{chu2019intelligent}} &
  \multirow{2}{2.2cm}{\centering MISO, Eve=K} &
  \multirow{2}{2.5cm}{\centering Approximation (AO-SDR)} &
  \multirow{2}{3cm}{\centering Rayleigh flat fading} &
  $\begin{multlined}
  O(N_r^2\sqrt{N_r}\log(1/\epsilon)\\(N_r^3 + 1 + N_r) + \\
  N_r^2(N_r^2 + 1 + N_r)+N_r^4)
  \end{multlined}$ &
  \multirow{2}{*}{SR} \\ \hline
\end{tabular}\label{table1}
\end{table*}

\begin{table*}[htbp]
\renewcommand{\arraystretch}{1.2}
\centering
\begin{tabular}
{
|>{\centering\arraybackslash}m{0.8cm}
|>{\centering\arraybackslash}m{2.2cm}
|>{\centering\arraybackslash}m{2.5cm}
|>{\centering\arraybackslash}m{3cm}
|>{\centering\arraybackslash}m{4.8cm}
|>{\centering\arraybackslash}m{2.1cm}
|
}
\hline
\textbf{Ref.} & 
\textbf{System Model} & 
\textbf{Methodology} &
\textbf{Assumptions} &
\textbf{Complexity} &
\textbf{Metric} \\ \hline
\cite{Xu2019} &
  MISO, Bob=L &
  Approximation (AO-SDR-MO) &
  Interference cancellation at Eve &
  -- &
  Average Sum SR \\ \hline 
\cite{Yu2019} &
  MISO, Eve=K, Bob=L &
  Approximation (AO-SCA-SDR) &
  Interference cancellation at Eve \& Multiple RISs &
  $\begin{gathered}
  O(\log(1/\epsilon)((\sqrt{N_a}+\sqrt{N_e})K^3L^3 \\
  + (N_a^{5/2} + N_r^{5/2})K^2L^2) 
  \end{gathered}$ 
    &
  Average Sum SR \& SOP\\ \hline
  \cite{yang2020deep} &
  MISO, Eve=K, Bob=L &
  Approximation (Deep RL PDS-PER) &
  Outdated/real-time CSI, Rayleigh flat fading &
  -- &
  SR \\ \hline
\cite{chen2019intelligent} &
  MISO, Eve=K, Bob=L &
  Approximation (AO) &
  Discrete/continuous phases \& active Eves &
  $\begin{gathered}
  O((LK+1)K^2N_a^2)\\
  O((LK+N_r+1)(N_r+1)^2)
  \end{gathered}$ &
  Max min SR \\ \hline
\end{tabular}
\end{table*}

\begin{table*}[!t] \renewcommand{\arraystretch}{1.60}
\centering
\caption{Summary of methodologies and key assumptions in the literature}
\begin{tabular}{
|m{8.8cm}
|>{\centering\arraybackslash}m{2.5cm}
|>{\centering\arraybackslash}m{3.5cm}
|>{\centering\arraybackslash}m{1.5cm}
|}
\hline
\textbf{Methodology/System Model}              & \textbf{SISO}                        & \textbf{MISO}                                                                                        & \textbf{MIMO}                                 \\ \hline
{Machine learning (DNN, Deep RL)}  &    & \cite{song2020truly,yang2020deep} &     \\ \hline
{Jamming (Artificial Noise)}            &                                      & \cite{Wang2019,Guan2020,Xu2019}                                                                      & \cite{Hong2020}                               \\ \hline
{Analytical}                            & \cite{Yang2020,makarfi2019physical,makarfi2020reconfigurable,hong2020secrecy} &                                                                                                      &                                               \\ \hline
{Alternating Optimization (AO)}         &                   \cite{long2020reflections}                   & \cite{chensecrecy,Wang2019,shi2019enhanced,Guan2020,Xu2019,Yu2019,Shen2019,Cui2019,Yu,chu2019intelligent,chen2019intelligent,ning2019improving,xiao2020intelligent} & \cite{Dong2020,jiang2020intelligent,alexandropoulos2020safeguarding}          \\ \hline
{Minorization Maximization/Majorization Minorization (MM)}        &                                      & \cite{shi2019enhanced,Yu}                                                                                            & \cite{Hong2020}                               \\ \hline
{Path-Following}                        &                                      & \cite{chen2019intelligent}                                                                                  &                                               \\ \hline
{Semidefinite Relaxation (SDR)}         &                                      & \cite{Feng2019,Wang2019,shi2019enhanced,Xu2019,Yu2019,Cui2019,chu2019intelligent}                                    &                                               \\ \hline
{Successive Convex Approximation (SCA)} &    \cite{long2020reflections}                                  &    \cite{shi2019enhanced,Yu2019}                                                                                                  & \cite{jiang2020intelligent}                   \\ \hline
{Projected Gradient Descent/Ascent (PGD/PGA)}      &                                      & \cite{Feng2019}                                                                                      &   \cite{alexandropoulos2020safeguarding}                                            \\ \hline
{Block Coordinate Descent (BCD)}        &                                      & \cite{Yu}                                                                                            & \cite{Hong2020}                               \\ \hline
{RIS Quantized Phase Shifts}            &                                      & \cite{chensecrecy,chen2019intelligent}                                                                           & \cite{jiang2020intelligent}                   \\ \hline
{RIS Continuous Phase Shifts}           & \cite{Yang2020,makarfi2019physical,makarfi2020reconfigurable,long2020reflections} & \cite{Feng2019,Wang2019,Xu2019,shi2019enhanced,Shen2019,Yu2019,Cui2019,Guan2020,Yu,chu2019intelligent,ning2019improving,song2020truly,yang2020deep}          & \cite{Hong2020,Dong2020,jiang2020intelligent,alexandropoulos2020safeguarding} \\ \hline
{Manifold Optimization (MO)}                 &                                      & \cite{Xu2019,wang2020intelligent}                                                                                        &                                               \\ \hline
{Fractional Optimization (FO)}               &                                      & \cite{Wang2019,Shen2019,ning2019improving}                                                                                      &                                               \\ \hline
\end{tabular}\label{table2}
\end{table*}

\section{Challenges, Recommendations, and Future Research Directions} \label{sec4}
\begin{figure*}[!hb]
\centering
\includegraphics[trim={0.6in 0in 0.6in 0in},clip,width=1\linewidth]{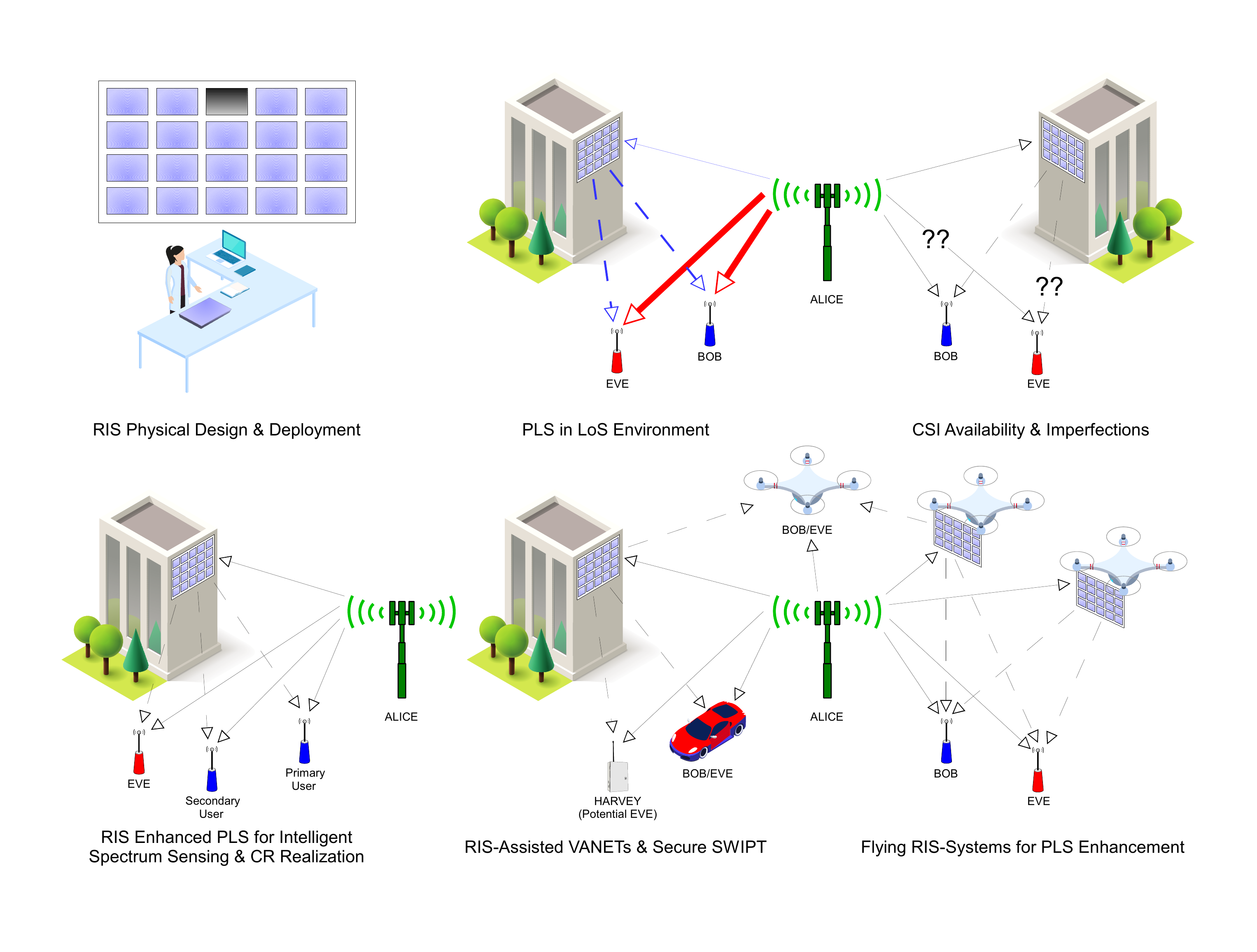}
\caption{Challenges and Future Directions}
\label{fig:challenges_and_future_directions} 
\end{figure*}

This section presents the challenges, recommendations, and future research directions for designing practical, efficient, and secure RIS-assisted future wireless communication systems, as summarized in  Fig.~\ref{fig:challenges_and_future_directions}. The conducted survey reveals that the simultaneous control of transmission from the BS and the reflections at the RISs can be an efficient solution to ensure confidentiality in wireless communication. Several simulation results verify the enhancement of the overall SR in such systems compared to the conventional ones. However, we stumbled upon several challenges and open directions for further investigation which we discuss along with recommendations, and future directions as follows.

\subsection{Effect of RIS physical design and deployment}
The effect of the physical design and deployment of RISs on PLS can be an interesting research direction, yet, it is not explored well in the literature. The physical design includes the number of RISs, their distribution, orientation, size, and geometrical shape. Moreover, the effect of varying the RIS's number of elements and their distribution on PLS needs further exploration. The effect of mobility and trajectory design in the case of mobile/flying RISs is yet to be studied, and the feasibility of using them in such scenarios is still an open problem.

Generally speaking, the deployment of RISs at different locations is a different problem compared to BSs/relays deployment because of the passive nature of RISs. Moreover, RISs are easier to deploy practically without interfering with each other due to their much shorter range when compared with active BSs/relays. However, how to optimally adjust the physical design, deployment, collaboration, and association to enhance RIS-assisted PLS is still an open challenge. ML and stochastic geometry-based solutions can be good options for efficient deployment of RIS assisted systems.

Furthermore, RIS-reinforced secrecy with imperfect RIS reflecting elements, i.e., discrete phase shifts and non-unit modulus (attenuating reflecting elements) also need to be considered while designing RIS-assisted PLS techniques.

\subsection{PLS in LoS Environments}
Ensuring confidential communication in the case of LoS transmission scenarios, where the eavesdropper is located within the same direction as that of the legitimate user, is quite challenging. Under these cases, several PLS techniques, including conventional beamforming, AN-based MIMO techniques, etc., \cite{8509094} will fail to provide secure communication. RIS can ensure secure communication even in such scenarios by providing additional channel paths between the legitimate nodes. There are few works reported in this direction \cite{Cui2019}, but further investigation is required, especially, when combining those scenarios with the imperfect or partially available CSI practical assumption.

\subsection{Effect of CSI availability and imperfect CSI}
{Based on the conducted survey, it is observed that the majority of the RIS-assisted PLS techniques in the literature assume the availability of perfect CSI at the transmitter and/or the RIS. However, channel estimation for the RIS assisted system is a challenging task due to a large number of passive reflecting elements. Moreover, these elements are passive in nature without signal processing capabilities and active transmitting/receiving abilities. Thus, in practice, only imperfect CSI can be accessed by the transmitter. Another issue that needs to be considered is that the CSI of an illegitimate node is available only if it is active or it is a licensed user that has legal access to the network. However, in the case of a passive eavesdropper, the CSI of the eavesdropper is not available. Moreover, in PLS literature, channel reciprocity property in TDD is assumed. However, with the involvement of RIS, this assumption may no longer be valid, which further complicates the problem \cite{wu2020intelligent}. A practical approach for acquiring the CSI at the RIS is proposed in \cite{taha2019deep,taha2019enabling}, where some of the reflecting elements are assumed to be active and can estimate the CSI, then, using compressive sensing or deep learning techniques, the CSI at all reflecting elements can be recovered/estimated. However, the channel is assumed to be sparse in the compressive sensing technique and requires a higher number of active elements as compared to the deep learning approach. 
Consequently, this calls for taking into account the effect of imperfect/incomplete CSI and its availability at transmitter while designing RIS-assisted security techniques for different channel models \cite{basar2020indoor} to ensure that these techniques are robust to these imperfections \cite{wu2020intelligent}}.

\subsection{Practical Realization and Higher-Order Metrics Assessment}
To show the effect of RIS-assisted secure communication in real environments, experimental work needs to be done. Although some promising experimental works have been reported in \cite{arun2019rfocus, liaskos2019novel, kaina2014shaping} to verify the gains offered by the RIS system, there is still a paucity of practical work for RIS-assisted secure communication. Moreover, the current practical work on RIS is not enough to decide the actual potential of RISs in practical conditions.

On the other hand, most of the current work focus on first-order metrics for PLS assessment {such as SR/SC and SOP. In this review, we highlighted new emerging higher-order metrics, which can add more insights to enable the actual realization of practical RISs. When security is a concern in adaptive transmission schemes and dynamic deployment of systems, ASOR and ASOD metrics can help in both their design and deployment. ASOR helps in quantifying the average secrecy level crossing rate at a specific secrecy threshold level predefined in the system, which in a sense shows ``how many times'' the communication system was vulnerable. On the other hand, ASOD specifies for ``how long'' this vulnerability was attained. Moreover, ASL, which is based on SR statistics, is also highlighted to quantify the amount of information that was lost in vulnerability durations \cite{li2020amount}.}

\subsection{Flying RIS-Systems for PLS Enhancement}
Recently, flying RISs-assisted communications (UAVs equipped with an RIS)  have received much attention. Some key features of flying RISs include 3D mobility, changeable direction and location, easy deployment, adaptive altitude, and power-efficient beamforming \cite{zhang2019reflections,8883124}. The trajectory of a flying RIS can be optimized along with the phase shifts adjustment at the RIS elements for enhancing PLS \cite{long2020reflections}. More specifically, the positioning/trajectory of the UAVs can be adjusted more flexibly in 3D space compared to terrestrial RISs. This feature can be used to improve the overall security by adapting the transmission based on the requirements, location, and channel conditions of the legitimate user. Besides, flying RISs can also be used as mobile cooperative jammers jointly with active UAVs or ground BSs to improve the secrecy performance. Moreover, in practice, a single flying RIS has limited capabilities in terms of communication and maneuvering. Hence, in some challenging scenarios, it may not achieve the desired secure communication performance, which motivates the investigations on multiple flying RISs along with active UAVs.

\subsection{Integration of RISs with Emerging Technologies and Future Applications}
RIS-assisted PLS solutions against passive and active eavesdropping for emerging and state-of-the-art technologies comes naturally. Promising research directions are, but not limited to, millimeter-wave communications, ma-MIMO, visible light communications, drones-aided communications, internet of things (IoT), THz communication, free-space optics, full-duplex communication, non-orthogonal multiple access (NOMA) \cite{furqan2019physical} and VANET \cite{8796365}.

Moreover, designing effective, adaptive, and intelligent \cite{yilmaz2017cognitive,2019arXiv190505075F} RIS-assisted PLS techniques under joint consideration of security, reliability, latency, complexity, and throughput based on QoS requirements of future applications to support URLLC, eMBB, and mMTC is also an interesting area of research. Furthermore, RIS-assisted cross-layer security design including the interaction of different layers, such as the physical layer, media access control (MAC) layer, network layer, and application layer, is not yet studied in the literature from the physical layer perspective.

\subsection{Optimization Problems to Enhance PLS}
Although the adoption of RISs in communication systems can enhance the overall systems' security, it results in higher complexity systems in terms of design and analysis as compared to conventional wireless systems \cite{8543573}. The use of data-driven tools such as ML, deep learning, and RL, is a promising solution to support the flexibility and the self-optimizability of such networks. Only a few works, \cite{song2020truly,yang2020deep}, have reported employing ML-based approaches to solve the PLS problem in RIS-assisted networks.

\subsection{RIS-Assisted Secure SWIPT}
{SWIPT is a promising technology to power massive low-power devices in the IoT for future wireless networks. The employment of RIS can enhance the performance of both the received information as well as the received energy in SWIPT systems \cite{wu2020intelligent}. In \cite{shi2019enhanced}, a system with an access point, information eavesdropper, legitimate information receiver, and a separated legitimate energy harvesting receiver, is investigated. It is reported that with the support of an RIS, the harvested power can be significantly improved under secrecy constraints. However, an untrusted energy harvesting receiver (HARVEY) can also eavesdrop the information intended for legitimate receivers. Designing efficient RIS-assisted secure SWIPT to prevent the untrusted energy harvesting receiver from eavesdropping the information is an interesting area for future research consideration. Obviously, this problem should be investigated under practical assumptions such as discrete phase shifts at the RIS, the coupling between the RIS elements' phase and the reflection gain, and the imperfect/unavailable eavesdropper's CSI}.

\subsection{RIS Enhanced PLS for Intelligent Spectrum Sensing and CR Realization}    
Intelligent spectrum sensing involves user detection, interference identification, and resource prediction, where those processes are often needed to be done in a secure manner. RISs can be utilized to create a secure environment against eavesdroppers in which spectrum sensing can be conducted reliably and securely. Moreover, RISs can be enablers for realizing secure CR systems, where RISs are used to ensure secured communication in a targeted network (primary or secondary).
\section{Conclusions}\label{sec5}
PLS supports the transmission secrecy when the conventional cryptographic methods fail due to the limited computational capacity at the legitimate communicating pairs or due to computationally over-powered eavesdropper. Yet, the efficiency of PLS is limited in some scenarios, for instance in the case of highly correlated legitimate and eavesdropping channels. RISs can be looked at as a promising solution in such scenarios, not to mention others, in the sense of adding more degrees of freedom by involving the propagation channel manipulation into the design problem. The most common, in addition to the recently proposed, performance metrics in PLS analysis were discussed. Furthermore, the PLS related works under the RIS-assisted networks have been reported and classified based on the adopted system model and the adopted methodologies.

Insightful recommendations are revealed upon this survey regarding the availability of the CSI, RIS design and deployment challenges, and the ML-based approaches to tackle the computational complexity encountered in all surveyed works. The deployment and orientation of RISs are key factors in reaping their full benefits in terms of system secrecy level. Hence, flying RISs are identified as a promising research direction, as they add more flexibility in the network by optimizing the RISs' 3D location, orientation, and trajectory to boost the system secrecy as well as to improve the overall energy efficiency.

\end{document}